\documentclass[%
 reprint,
 superscriptaddress,
 amsmath,amssymb,
 aps,
]{revtex4-2}

\usepackage{graphicx}
\usepackage{dcolumn}
\usepackage{bm}
\usepackage{xcolor} 

\begin{document}

\preprint{APS/123-QED}

\title{Large-scale error-tolerant programmable interferometer fabricated by femtosecond laser writing}

\author{I.V. Kondratyev}
\affiliation{%
 Quantum Technology Centre and Faculty of Physics, M.V. Lomonosov Moscow State University, 1 Leninskie Gory Street, Moscow 119991, Russian Federation
}%
\email{iv.kondratjev@physics.msu.ru}
\author{V.V. Ivanova}
\affiliation{%
 Russian Quantum Center, Bolshoy bul'var 30 building 1, Moscow 121205, Russian Federation
}%
\author{S.A. Zhuravitskii}
\affiliation{%
 Quantum Technology Centre and Faculty of Physics, M.V. Lomonosov Moscow State University, 1 Leninskie Gory Street, Moscow 119991, Russian Federation
}%
\author{A.S. Argenchiev}
\affiliation{%
 Quantum Technology Centre and Faculty of Physics, M.V. Lomonosov Moscow State University, 1 Leninskie Gory Street, Moscow 119991, Russian Federation
}%
\author{N.N. Skryabin}
\affiliation{%
 Quantum Technology Centre and Faculty of Physics, M.V. Lomonosov Moscow State University, 1 Leninskie Gory Street, Moscow 119991, Russian Federation
}%
\author{ \\ I.V. Dyakonov}
\affiliation{%
 Quantum Technology Centre and Faculty of Physics, M.V. Lomonosov Moscow State University, 1 Leninskie Gory Street, Moscow 119991, Russian Federation
}%
\affiliation{%
 Russian Quantum Center, Bolshoy bul'var 30 building 1, Moscow 121205, Russian Federation
}%

\author{S.A. Fldzhyan}
\affiliation{%
 Quantum Technology Centre and Faculty of Physics, M.V. Lomonosov Moscow State University, 1 Leninskie Gory Street, Moscow 119991, Russian Federation
}%
\affiliation{%
 Russian Quantum Center, Bolshoy bul'var 30 building 1, Moscow 121205, Russian Federation
}%

\author{M.Yu. Saygin}
\affiliation{%
 Quantum Technology Centre and Faculty of Physics, M.V. Lomonosov Moscow State University, 1 Leninskie Gory Street, Moscow 119991, Russian Federation
}%
\affiliation{Laboratory of Quantum Engineering of Light, South Ural State University (SUSU), Russia, Chelyabinsk, 454080, Prospekt Lenina 76}%

\author{S.S. Straupe}
\affiliation{%
 Quantum Technology Centre and Faculty of Physics, M.V. Lomonosov Moscow State University, 1 Leninskie Gory Street, Moscow 119991, Russian Federation
}%
\affiliation{%
 Russian Quantum Center, Bolshoy bul'var 30 building 1, Moscow 121205, Russian Federation
}%
\author{ \\ A.A. Korneev}
\affiliation{%
 Quantum Technology Centre and Faculty of Physics, M.V. Lomonosov Moscow State University, 1 Leninskie Gory Street, Moscow 119991, Russian Federation
}%
\author{S.P. Kulik}
\affiliation{%
 Quantum Technology Centre and Faculty of Physics, M.V. Lomonosov Moscow State University, 1 Leninskie Gory Street, Moscow 119991, Russian Federation
}%
\affiliation{Laboratory of Quantum Engineering of Light, South Ural State University (SUSU), Russia, Chelyabinsk, 454080, Prospekt Lenina 76}

\date{\today}

\begin{abstract}
We introduce a programmable 8-port interferometer with the recently proposed error-tolerant architecture capable of performing a broad class of transformations. The interferometer has been fabricated with femtosecond laser writing and it is the largest programmable interferometer of this kind to date. We have demonstrated its advantageous error tolerance by showing an operation in a broad wavelength range from $920$ to $980$ nm, which is particularly relevant for quantum photonics due to efficient photon sources. Our work highlights the importance of developing novel architectures of programmable photonics for information processing.

\end{abstract}

\maketitle


\section{\label{sec:introduction}Introduction}

Programmable multiport interferometers (PMI) are targeted at precise, energy-efficient and compact manipulation of information encoded in multiple modes of optical fields~ \cite{Harris18, Perez2018}. The growing interest in PMIs is fueled by a significant number of applications: optical switching in telecommunications \cite{Tanizawa15, Lu16, Suzuki17}, matrix-vector multiplication in optical neural networks \cite{Zhou2022, Cheng2022} and quantum information processing \cite{Capmany2020, Carolan2015, sibson2017chip}. Broadband operation \cite{luo2019design} and low power consumption \cite{goel2020survey} are two special features that are driving the growing interest of the scientific community and industry in information processing with PMIs.

A PMI is a static waveguide structure endowed with tunable phaseshifters (PSs). In $N$-port PMIs, a phase pattern induced by a set of PSs programs a specific $N \times N$ unitary transformation matrix $U$, which should be applied  to an input vector of light field amplitudes. The PMI architecture can be either dictated by a particular task \cite{sibson2017chip, peruzzo2014variational} or represent a general class of PMIs often called universal photonic processors. A universal photonic processor provides access to a full space of unitary transformations of input light vectors and the transformation $U(\{\phi\})$ is precisely controlled by phases $\{\phi\}$. Conventional universal photonic processor architectures rely on a well-known unitary matrix decomposition into a sequence of two-dimensional subspace rotations \cite{Reck1994, Clements16}. Each subspace rotation can be implemented using a Mach-Zehnder interferometer (MZI) acting on a particular pair of modes belonging to a chosen subspace. Each MZI matrix $U_{MZI}(\theta,\phi)$ is essentially a $2\times 2$ variable beamsplitter matrix embedded in a larger $N\times N$ unitary of a complete PMI. Such PMI architectures are particularly appealing due to the existence of an analytical algorithm that computes phase sets $\{\theta\}$ and $\{\phi\}$ corresponding to a desired PMI unitary matrix $U$. The universality of these architectures hinges on the ability of an ideal MZI to cover the entire $SU(2)$ group, which implies that the static beamsplitters (BSs) constituent the MZIs must be ideally balanced. However, non-perfect manufacturing quality of the static elements hinders the universality of the fabricated PMIs. State-of-the-art technology and design tools deal with the majority of problems providing wafer-scale fabrication of high-quality components \cite{kalaiselvi2022wafer} which are engineered to be robust \cite{Saygin2020}. However, errors still may creep once a truly large-scale interferometers are fabricated \cite{bao2023very} and may even evolve during long-term operation. 

\begin{figure*}[ht!]
\centering
\includegraphics[width=2\columnwidth]{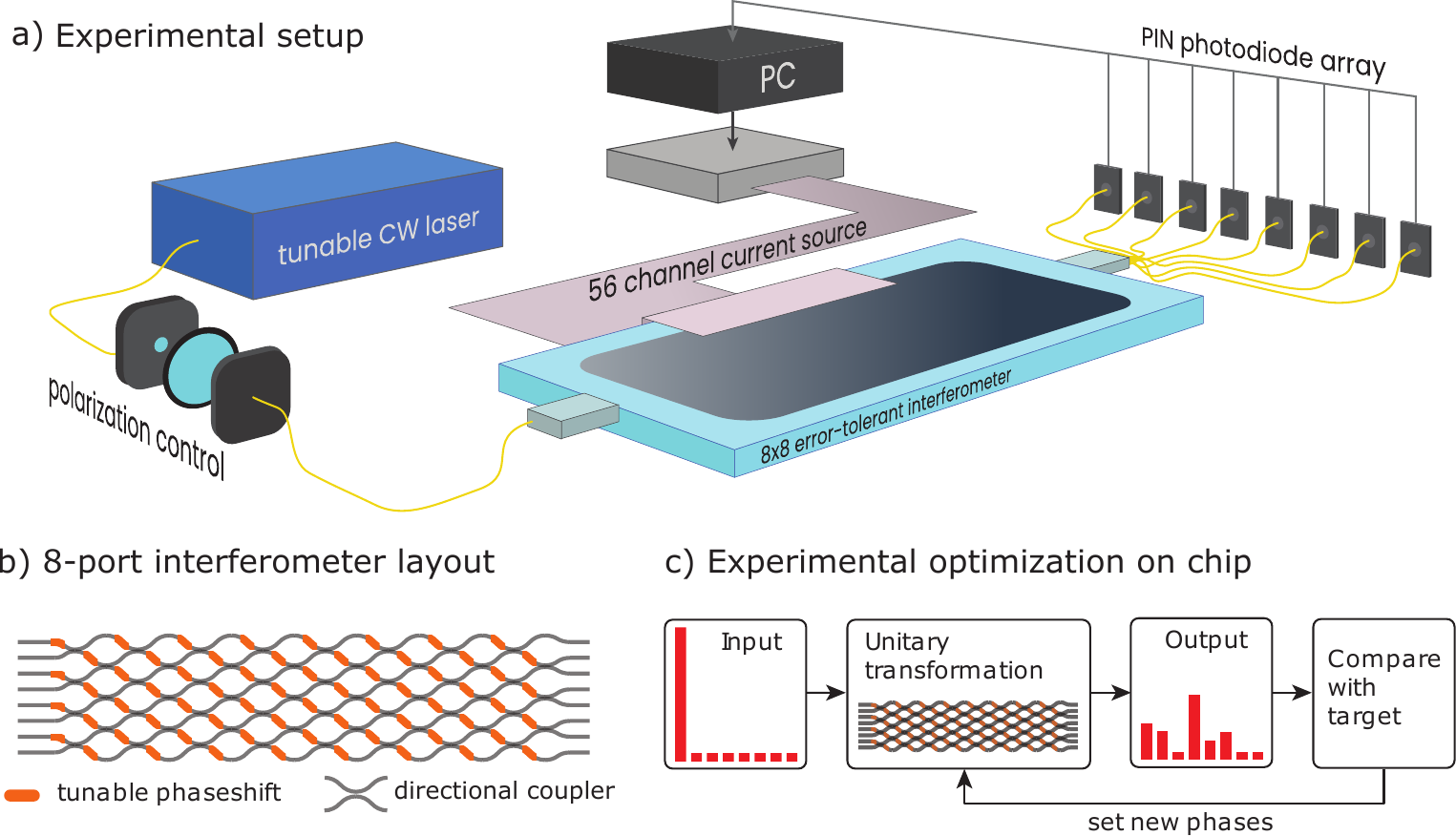}
\caption{\label{fig::PMI_layout}a) Sketch of the experimental setup. The waveguide structure of the 8-port reconfigurable interferometer consists of 56 directional couplers and 56 PSs. The optical chip was connected to a 64 channel current source capable of setting currents up to 60 mA in each of its channels individually with a step of 0.01~mA. The current source was connected to and fully controlled from a PC. b) Scheme of an eight-mode error-tolerant interferometer architecture. Each of the 56 directional couplers has an imbalanced splitting ratio shifted to a higher transmission according to original proposal \cite{Fldzhyan20}. c) Illustration of the optimization principle used for programming the interferometer. }
\end{figure*}

Therefore, developing more advanced interferometer architectures that would tolerate high levels of static errors is paramount and could facilitate the creation of sophisticated photonic processors. Several works have studied usage of alternative building blocks as a foundation of robust architectures instead of MZIs, in particular, BSs and multiport couplers~\cite{Fldzhyan20, Saygin2020}. It turns out that specific arrangements of these devices and phaseshifting elements can lead to architectures that tolerate large deviations of their static elements from ideal counterparts, which relaxes the requirements to fabrication tolerances and expands the range of possible applications. For instance, since the unitary matrix of both the BS and the multiport coupler is susceptible to a signal wavelength, the PMI with a robust architecture can work as a more broadband optical processor than the MZI-based one, which, in its turn, may also support broadband operation by careful engineering of its static components \cite{burgwal2017using}, however this step usually complicates both development and fabrication processes and results in increasing the overall size of the interferometer \cite{miller2015perfect,burgwal2017using}.

The state-of-the-art PMI fabrication technology is lithography-based process \cite{Silverstone16} which offers compatibility with standard CMOS production line \cite{siew2021review}. This approach provides both flexible and extremely precise tools for the fabrication of nano- and microscale photonic structures. The largest PMIs to date were fabricated using lithography-based Triplex technology \cite{taballione2021universal}. The quality of the fabricated PMIs can be estimated either by comparing an implemented unitary transformation with a desired counterpart \cite{laing2012super}, which is a tedious procedure, or by running a specific task and checking the device performance. For instance, a deep interest in PMIs grows from their natural ability to multiply an input complex field vector by a unitary matrix. Hence the quality may be assessed by checking the precision of this operation. This method has found wider application in PMI quality estimation \cite{tang2021ten}. Tests using both methods indicate that lithographically fabricated processors can be of very high quality \cite{Carolan2015}. In this work we use an alternative fabrication technology~---~femtosecond laser writing (FSLW)~---~that can be used for rapid and cost-effective PMI prototyping \cite{Cai22}. It has been widely adopted by the scientific community not only due to its relative simplicity and potential to produce PMIs with competitive quality, but also since it has a feature to fabricate waveguide structures in three dimensions in comparison with the lithography-based process whose waveguide geometries are inherently planar \cite{flamini2015thermally, ceccarelli2020low, skryabin2023two}. Recently, a 6-port universal MZI-based PMI fabricated by FSLW with 30 thermo-optical modulators was reported \cite{pentangelo2022universal}. On the other hand, the FSLW technology offers less precision and PMIs suffer from substantially larger cross-talks between PSs \cite{ceccarelli2019thermal}. Therefore, this technology should benefit especially from robust architectures. Regardless of the technological platform, robust architecture can significantly improve PMI quality, making it a universal method for error-prone optical multiport design.

In this paper, we demonstrate an 8-port PMI based on the error-tolerant architecture \cite{Fldzhyan20}. The interferometer is fabricated using FSLW technology and includes 56 directional couplers (DCs) and 56 thermo-optical PSs. To the best of our knowledge, this is the largest PMI that has been fabricated using FSLW to date. We demonstrate broadband port-to-port switching by tuning the interferometer with an optimization procedure.
The reported result indicates that the error-tolerant architecture adds robustness even to PMIs which are fabricated with less stringent technological process.

\begin{figure}[ht!]
\centering
\includegraphics[width=1.\columnwidth]{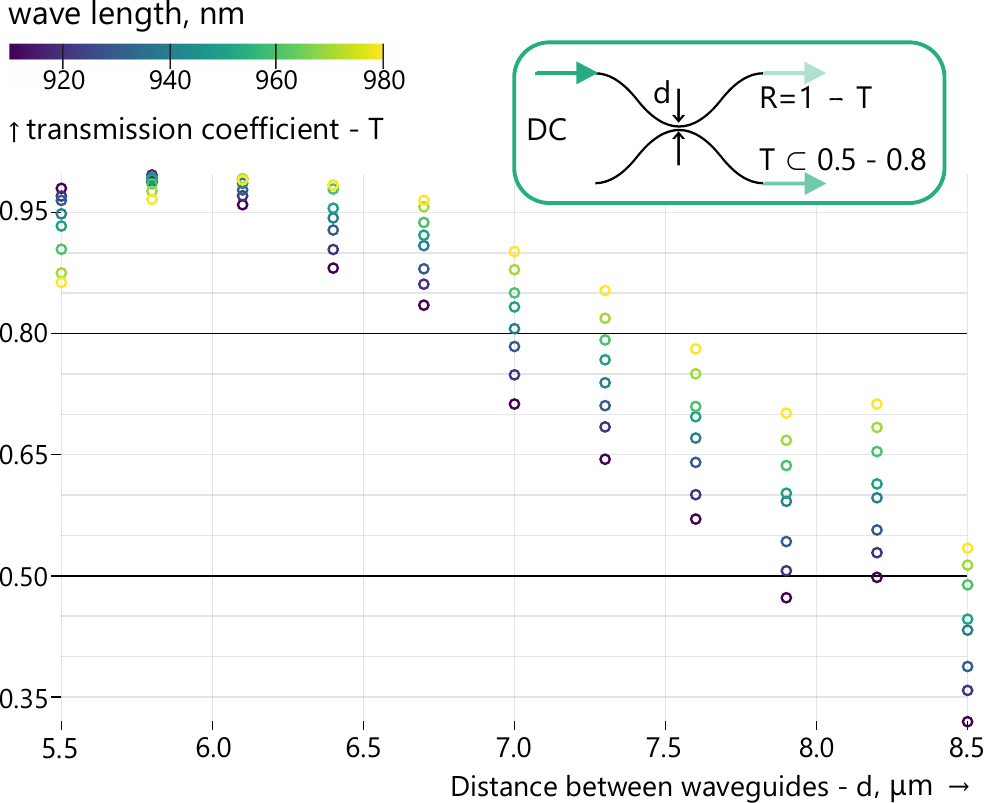}
\caption{\label{fig::T_for_11_d} The dependence of the transmission coefficients on the distance between the waveguides in the DC at different wavelengths. The green dotted lines limit the transmittance range from 0.5 to 0.8 which is required by the PMI architecture \cite{Fldzhyan20}. Inset schematically shows the DC structure. }
\end{figure}

\section{\label{sec:fab_and_exp} Experimental setup and interferometer fabrication} 

The general experiment setup is sketched in Fig. \ref{fig::PMI_layout} (a) having in its core the 8-port PMI, which was fabricated by the FSLW technique in fused silica glass (JGS1, AGOptics) sample with the size of 100x50x5~mm$^3$ (see Appendix \ref{app:fab_process} for details). The waveguide architecture of the studied PMI is shown in Fig. \ref{fig::PMI_layout} (b) with the input and output ports to be separated by a 127~$\mu$m gap to interface with v-groove single-mode fiber arrays. In the measurement setup input and output fiber arrays are connected to the tunable CW diode laser (Toptica CTL 950) and  eight photodetectors, respectively. The polarization of the input radiation was controlled by the HWP and QWP plates installed in the free-space area before the input fiber array. The PSs are implemented as thin metallic wires which induce refractive index change due to the thermo-optical effect. The NiCr film with a thickness of about 0.2~$\mu$m is deposited on the top surface of the optical chip by magnetron sputtering process. The heating wires and contact pads are engraved by laser ablation using the same FSLW setup. The geometric parameters of the heating wires are adjusted such that their typical resistances are around $R_{h}=450$ $\Omega$ (see Appendix \ref{app:inf_design} for details). A home-built 12-bit digital constant current source powers the heaters, and a specific printed circuit board with spring-loaded connectors interfaces the PMI with the current source. The optical chip was mounted on a temperature stabilized aluminum platform with a temperature setpoint around 20~$^{\circ}$C.

\begin{figure}[ht!]
\centering
\includegraphics[width=1.\columnwidth]{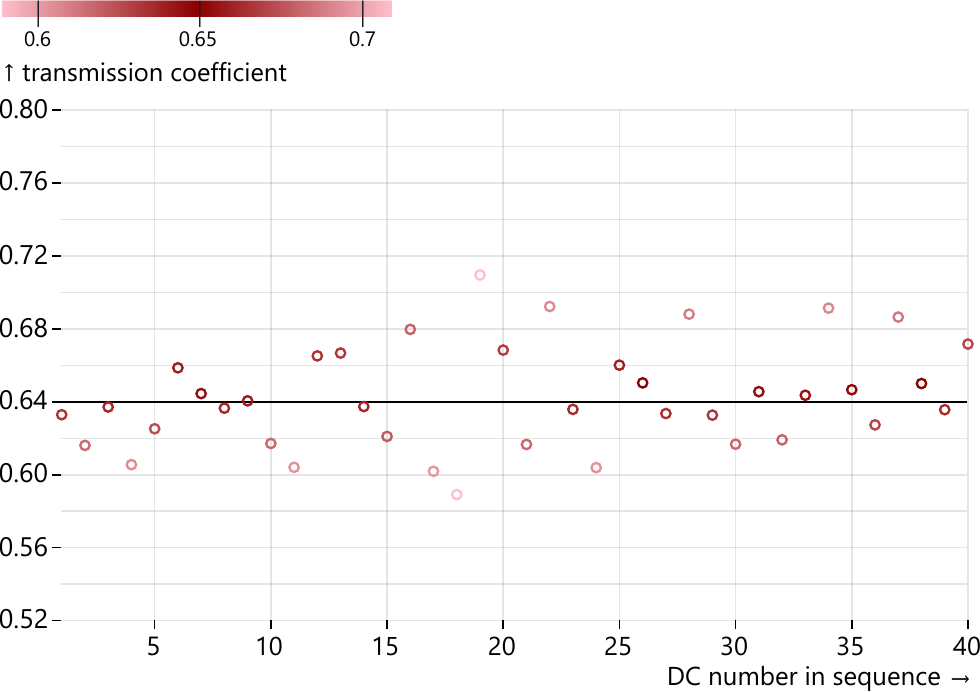}
\caption{\label{fig::forty_DC} Statistics of the transmission coefficient for 40 DCs with $d = 7.8$~$\mu$m at the 945~nm wavelength. Noticeable fluctuations in the absolute value of the transmission coefficient $T$ are clearly visible. However, $T$ falls in the range $[0.5, 0.8]$ for all 40 directional couplers.}
\end{figure}

As the key component of our PMI is a directional coupler (DC), its parameters were carefully calibrated. A generic DC is completely described with a $2\times 2$ unitary matrix $U_{DC}=\bigl( \begin{smallmatrix}\sqrt{1-T} & i\sqrt{T}\\ i\sqrt{T} & \sqrt{1-T}\end{smallmatrix}\bigr)$, where $T$ is the power transmission coefficient. The integrated photonic DC design includes a pair of waveguides with two circular arc s-bends with 60~mm radius and a minimum distance $d$ between the waveguide cores. In order to select the appropriate parameters of the DCs, a series of 11 DCs with varying distances $d$ were fabricated. The distance $d$ was set within the range of 5.5 - 8.5~$\mu$m with a step of 0.3~$\mu$m. The transmission coefficient $T$ was measured for each DC in the wavelength range from 910~nm to 980~nm with a step of 10~nm. The results are shown in Fig.~\ref{fig::T_for_11_d}. The error-tolerant interferometer architecture described in \cite{Fldzhyan20} requires the transmission coefficients of the DCs to be in the range from 0.5 to 0.8. The data in Fig.~\ref{fig::T_for_11_d} suggest that DCs with $7.5<d<8$~$\mu$m satisfy this requirement, and $d = 7.8$~$\mu$m was chosen for the fabricated PMIs.

\begin{figure*}[t!]
\centering
\includegraphics[width=2\columnwidth]{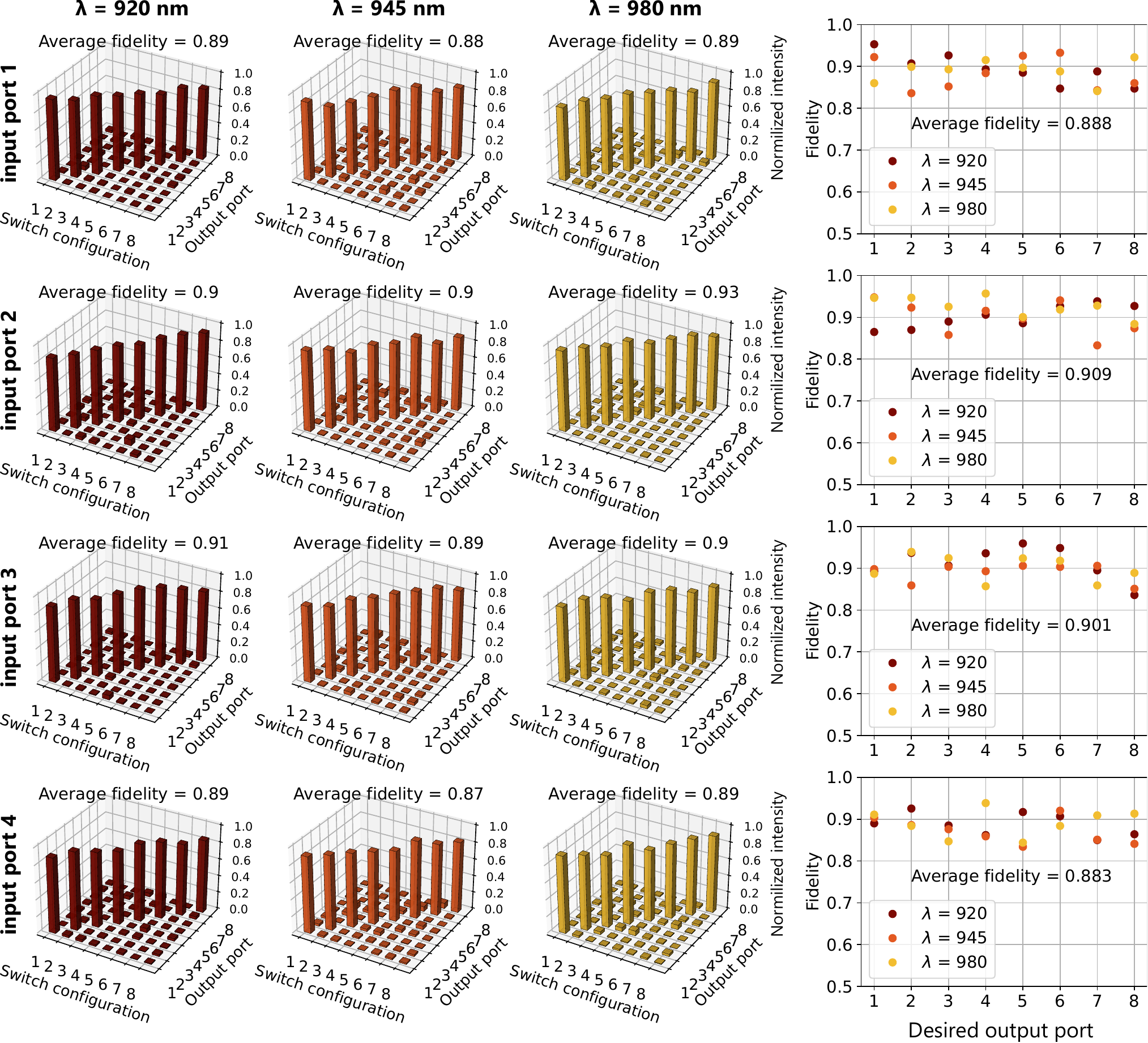}
\caption{\label{fig::permut_result} Results of VFSA optimisation of the phaseshifts to realize optical switching for three wavelengths: 920, 945 and 980 nm. Histograms of the power distribution in the output ports of the device optimized for switching to the specific output are shown for each input port and for each wavelength. The fidelity of the observed distribution to the expected one is shown for each input port in the graphs on the right.}
\end{figure*}

To verify the reproducibility of a single DC in our FSLW setup, 40 DCs with $d = 7.8$~$\mu$m were fabricated and the corresponding transmission coefficients $T$ were measured. The values obtained are shown in Fig.~\ref{fig::forty_DC}. It can be seen that, while values of $T$ underwent noticeable fluctuations, the transmission coefficients were still in the range $[0.5, 0.8]$ for all 40 DCs, which is necessary for the architecture of our interferometer to preserve its universality.

Finally, we fabricated the 8-port PMI comprised of 56  reconfigurable building blocks. Each block includes a DC with the $d=7.8$~$\mu$m gap between the waveguides in the interaction region and a thermo-optical PS. The blocks are arranged according to the error-tolerant architecture.
Registered time for complete reconfiguration of the chip, which included switching phases on all 56 PSs, was no longer than 2 seconds (see Appendix \ref{app:time_responses} for details).
We also observed thermal crosstalk between PSs in transverse direction in our PMI, which was  automatically taken into account during the optimization procedure ran on chip (see Appendix \ref{app:heater_crosstalk} for details).
Measured electrical power required for $2\pi$ phase switching on a single PS was 0.33 W (see Appendix \ref{app:power_consumed} for details).

\section{\label{sec:exp_and_results}Experiments and Results}

We tested the fabricated PMI by programming it to operate as an $8\times 8$ optical switch at three different wavelengths: 920, 945 and 980~nm. At each wavelength, the DC structures inside the PMI are characterized by a corresponding transmission coefficient (see Fig.~\ref{fig::T_for_11_d}), and thus we tested the performance of the error-tolerant architecture within the required transmission range $[0.5; 0.8]$. It is noteworthy that switching is a particularly difficult task for an MZI-based universal interferometer since perfect switching can only be achieved when the DC transmission $T$ is exactly $0.5$ or an optimization routine has to be introduced in order to minify the effect of non-perfect DC. We present a numerical comparison of switching performance of both error-tolerant and MZI-based architectures in Fig.~\ref{fig::comparison_permut}. The details of the simulation are provided in Appendix~\ref{app:comparison_clement}. It is clearly evident that the error-tolerant architecture is substantially more robust to the DC transmission coefficient variations. This means that the fabricated $8\times 8$ error-tolerant PMI demonstrates high-fidelity performance in a broad wavelength range 910-980~nm.

\begin{figure}[t!]
\centering
\includegraphics[width=\columnwidth]{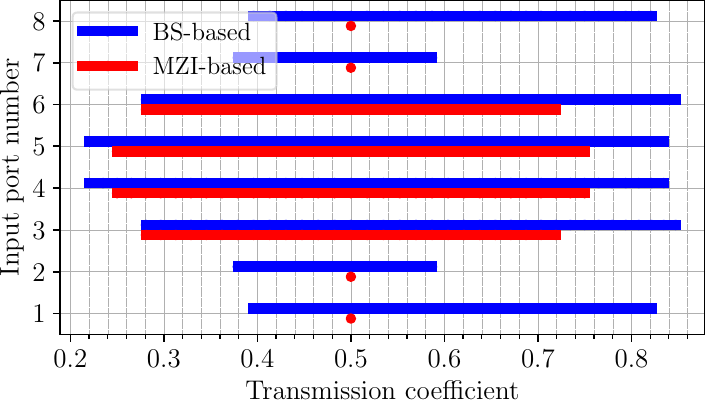}
\caption{\label{fig::comparison_permut} Simulated numerical comparison of the performance of realizing port-to-port optical mode switching between BS-based error-tolerant PMI \cite{Fldzhyan20} and conventional MZI-based PMI \cite{Clements16} architectures. The highlighted regions show the transmissions of DCs of permutation-capable PMI configurations.
}
\end{figure}

We experimentally program our PMI to implement switching configurations using a global optimization routine. The optimizer searches for the 56 phase values inside the interferometer that correspond to the desired output power distribution. We use a Very Fast Simulated Annealing (VFSA) algorithm \cite{vakil2008modified} to minimize the infidelity $\text{Inf}=1-F$ with the fidelity function defined as

\begin{equation}\label{eq::infidelity}
    F(X,Y)=\left(\sum_{j} \sqrt{ X_j Y_j }\right)^2,
\end{equation}

where $X$ and $Y$ are the normalized column vectors ($ \sum X_j = \sum Y_j  = 1$) corresponding to the experimentally measured and target output power distributions respectively. 

We conducted a series of optimization experiments by reconfiguring the chip to switch all optical power from the selected input port to each of the eight output ports as it is schematically shown in Fig. \ref{fig::PMI_layout} (c). The switching was performed using only the first four input ports. 
During the optimization process, the temperatures of the PSs were adjusted stepwise with a 10 second delay between the steps to maintain a constant chip temperature of 20 $^\circ$C.

The experimental results are shown in Fig.~\ref{fig::permut_result}. Each row of the bar charts illustrates the results of output power measurements from the optical chip optimized to realize a given configuration. The colors of the histograms encode the wavelength used in the experiment (red color corresponds to 920~nm, orange -- to 945~nm, yellow -- to 980~nm). The right column shows the fidelity values for each optimized output distribution after 500 iterations of the optimization algorithm. The convergence curves are provided in Appendix~\ref{app:converg}. The average switching fidelities for all input ports and wavelengths are in the range between 0.89 and 0.93. The results provide clear evidence that the PMI constructed using a BS-based robust architecture is capable of high-performance switching when the static DCs' transmissions lie in the $[0.5, 0.8]$ range. During the series of optimization experiments each of 56 PSs was switched no less than $48000$ times ended up undamaged, which evidences the high level of our PMI durability (see Appendix \ref{app:durability} for details).

\section{\label{sec:discussion}Conclusion} 

We have demonstrated the FSLW-fabricated PMI with the BS-based robust architecture by showing its broadband optical power switching operation. The wavelengths of 920, 945 and 980~nm were selected to highlight the robustness of the PMI architecture to BS reflectivity variations. Having in mind the application of the PMIs in quantum photonics, in particular, for processing high-dimensional quantum states the wavelength selection range is advocated by the recent introduction of quantum dot single photon sources to quantum photonics \cite{arakawa2020progress} that operate efficiently around 910-940~nm. Since the fabrication of a QD source with a fixed required emission wavelength is still a major technological challenge the demonstrated PMI can be considered as a reliable photonic platform compatible with the sources generating at very different wavelengths. In other words, with the robust architecture there is no need in customizing the PMI to a specific wavelength.

Even though we have only performed PMI programming in the classical regime, each optimization run still takes a few dozens of iterations and about an hour to converge properly. However, once the phase settings for each switching configuration have been established, the PMI does not need to be recalibrated. Unfortunately, the task of programming a PMI with a BS-based architecture, exact BS parameters of which are not precisely known, becomes a black box problem that has no simple analytical solution. Recently, methods were proposed to reconstruct the internal structure of an interferometer using auxiliary measurements \cite{kuzmin2021architecture, bantysh2023fast, maring2023general}. These methods apply well-known machine learning techniques to yield the unitary matrices of individual components in the photonic circuit of an interferometer and, as a result, to develop an accurate numerical model of a device under study. This model helps to transfer the optimization task from a real device to a corresponding numerical model, thus greatly simplifying the process of programming the required unitaries in PMIs with complex architectures.

It should be noted that there is room for improvement of the PMIs with our architecture using technological and layout improvements, which have been demonstrated recently. The thermooptical phasesifter power efficiency and the crosstalk can be improved by the orders of magnitude by augmenting the structure with the thermal insulation trenches \cite{ceccarelli2020low}. The slightly more complex FSLW waveguide fabrication process \cite{pentangelo2022universal} enables lower propagation loss and higher refractive index contrast which in turn positively affects the miniaturization and further upscaling of the FSLW reconfigurable photonics. This makes us believe that femtosecond direct laser writing will stay a technology of choice for rapid and affordable prototyping.

\section{\label{sec:acknowledgements}Acknowledgements}
The work was supported by Rosatom in the framework of the Roadmap for Quantum computing (Contract No. 868-1.3-15/15-2021 dated October 5, 2021 and Contract No.P2154 dated November 24, 2021) in part of fabrication of integrated photonic chips. The work was supported by Russian Science Foundation grant 22-12-00353 (https://rscf.ru/en/project/22-12-00353/) in part of the development and experimental assessment of the reconfiguration algorithm.

\newpage{}
\nocite{*}

\bibliography{apssamp}

\newpage{}

\appendix

\section{Fabrication process \label{app:fab_process}}

The waveguide circuit was written with a 515~nm laser pulses (second harmonic of an Avesta Antaus ytterbium fiber femtosecond laser system) with 280~fs duration delivered at 1~MHz repetition rate with 120~nJ pulse energy and linear polarization parallel to the writing direction inside a fused silica glass sample. The laser beam was focused with the aspheric lens (NA~=~0.55) 15~$\mu$m below the surface of the sample. A 150~$\mu$m thick cover glass was placed between the lens and the sample for partial correction of spherical aberrations. A reconfigurable beam expander was used to fill the input aperture of the focusing lens. A high-precision AeroTech FiberGlide3D air-bearing system was used to moving the sample at a speed of 0.2~mm/s during the waveguide fabrication process.
The inscribed waveguides have the refractive index contrast $dn \approx 10^{-3}$ and a slightly asymmetric $5\times 9$~$\mu$m nearly Gaussian eigenmode. The average propagation loss is 0.6~dB/cm~@~910~nm and the coupling loss is 1.5~dB per endface, the additional bending loss is \textless{0.1}~dB/cm for a 60~mm bending radius used in the experiment.

\begin{figure*}[t!]
\centering
\includegraphics[width=2.1\columnwidth]{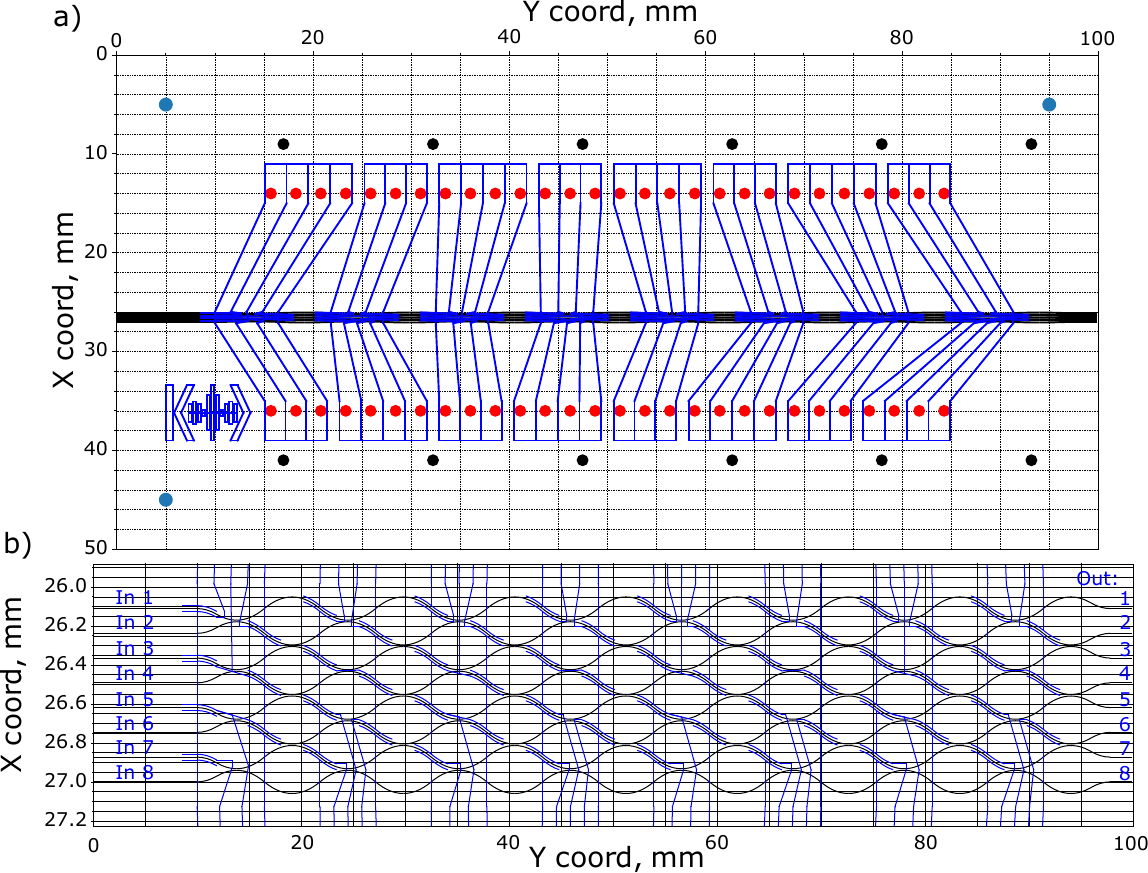}
\caption{\label{fig::waveguides_n_electrodes}Actual optical chip structure (view from the top). a) Real scale scheme of the chip. b) Zoomed part of the waveguide structure. Waveguides are depicted with black solid lines, engraved electrodes are depicted with blue solid lines. Red and black dots represent the electrical and ground contacts with the PCB.}
\end{figure*}

\section{Interferometer design \label{app:inf_design}}

The interferometer design, which includes the waveguide structure and the electrode pattern, is shown in Fig.~\ref{fig::waveguides_n_electrodes}. The radius of curvature is 60~mm for all bends, which is a trade-off between the bending loss effect and the overall size of the waveguide structure achievable using our fabrication technology. The input and output ports are spaces 127~$\mu$m apart and interfaced with single-mode fiber arrays. 
The thermooptical phaseshifters are 30~$\mu$m wide and 2.7~mm long metal stripes. The transverse (perpendicular to the waveguide axis) spacing between the heaters is 240~$\mu$m Fig.~\ref{fig::waveguides_n_electrodes}. The NiCr metallic film is 0.2~$\mu$m thick. The electrodes connect to a multi-channel digital computer controlled current source via a PCB interface with spring-loaded contacts. The resistances of all heaters fall within 400 - 500~$\Omega$ range.

\begin{figure*}[ht!]
\centering
\includegraphics[width=2\columnwidth]{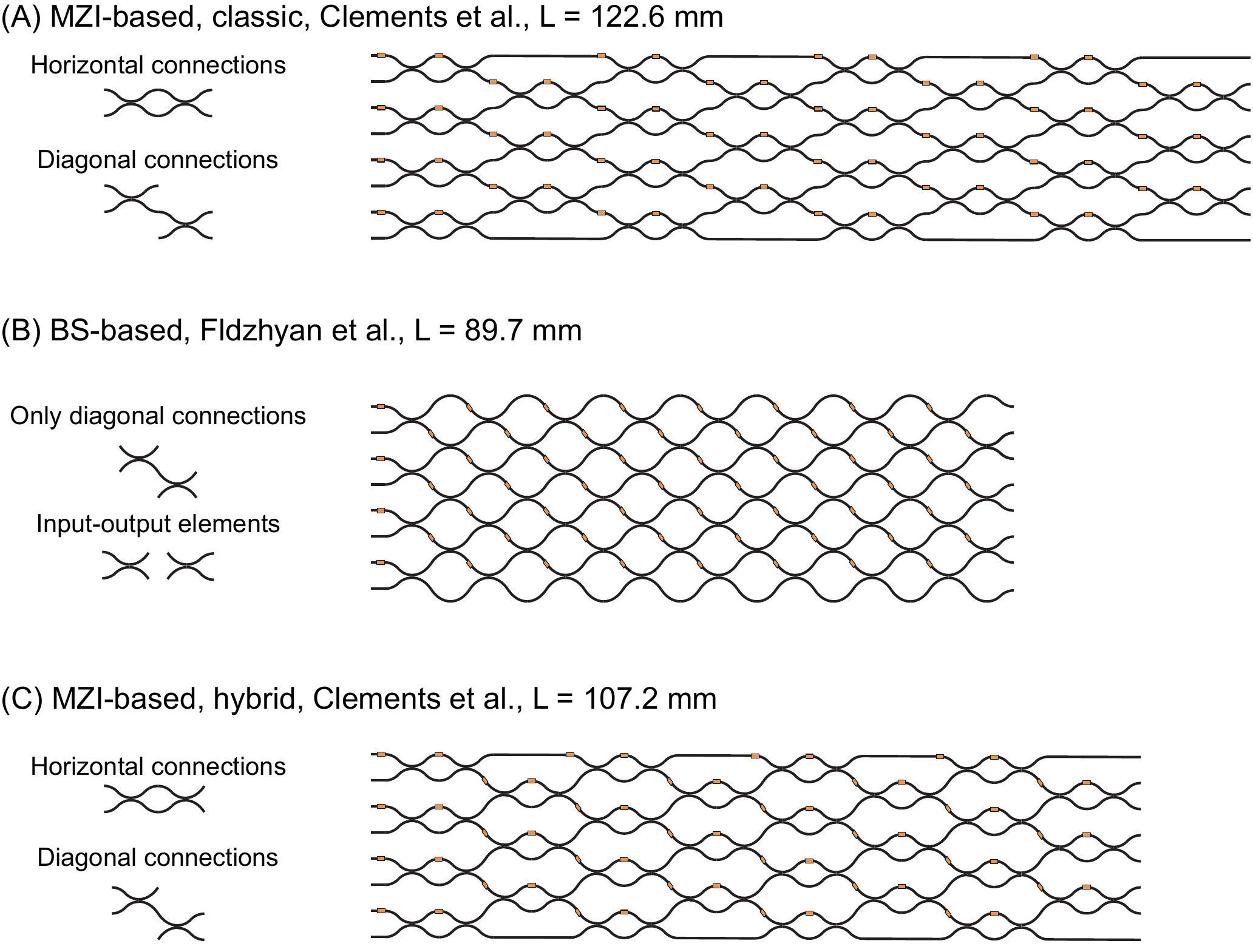}
\caption{\label{fig::architectures_comparison} Comparison of the total lengths of eight-mode interferometers with different architectures with a curvature radius of R = 60~mm and an input/output distance between ports equal to 127~$\mu$m. a) Scheme of a MZI-based interferometer \cite{Clements16} with straight waveguides connecting individual directional couplers. b) Scheme of the BS-based error-tolerant interferometer \cite{Fldzhyan20} demonstrated in this work including diagonally connected directional couplers. c) Scheme of a MZI-based interferometer with optimized directional coupler connections.}
\end{figure*}

We also compared the total lengths of eight-mode interferometers with different architectures with a same curvature radius of R = 60~mm and an 127~$\mu$m pitch between input/output ports (see Fig.~\ref{fig::architectures_comparison}).
With these parameters an MZI-based interferometer \cite{Clements16} which directional couplers are connected to each other with straight waveguide sections has a total circuit length of more than L~=~122~mm (see Fig.~\ref{fig::architectures_comparison}a). In contrast, a BS-based error-tolerant interferometer \cite{Fldzhyan20} fabricated in this work is shorter by more than 25\% and has the total circuit length of less than L = 90~mm (see Fig.~\ref{fig::architectures_comparison}b). This layout has diagonal connections between DCs, which helps to truncate the optical circuit. The error-tolerant interferometer has balanced propagation and bending losses throughout the interferometer. If the straight connections in the MZI-based interferometer are replaced with diagonal ones (see Fig.~\ref{fig::architectures_comparison}c) the total length shrinks to around 107~mm, which is, however, still more than 15\% longer than the BS-based error-tolerant PMI manufactured in this work.

\section{Times of reconfiguration \label{app:time_responses}}

\begin{figure*}[t!]
\centering
\includegraphics[width=2.1\columnwidth]{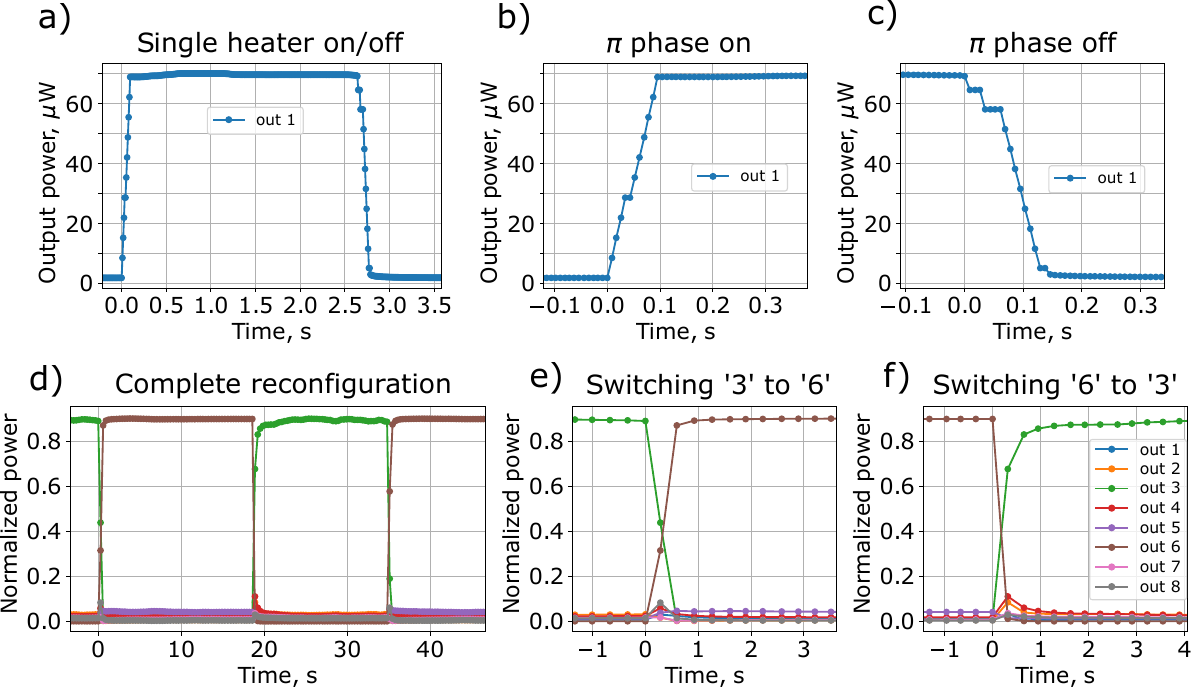}
\caption{\label{fig::time_responses}Measured times of reconfiguration of the fabricated optical chip: a) Turning on and off the phaseshift $\pi$ on a single heater. b) Applying phase $\pi$  on a single heater takes $\approx$ 100 ms c). Turning the phase $\pi$ off on a single heater lasts no longer than 150 ms d). Switching the chips' transformation from configuration where most of output power exits the third output port to configuration where most of the power exits from the sixth output port and back. e) Switching from the third output to the sixth output takes $\approx 1$~s. f) Switching from the sixth output to the third output takes $\approx 2$~s. }
\end{figure*}

We measured typical thermo-optical heater time responses both by applying a $\pi$ phase on a single heater and by registering the time of complete reconfiguration of the chips' transformation, which implies changing all 56 phaseshifts at a time. The results are shown in Fig. \ref{fig::time_responses}. It can be seen that applying a phaseshift on a single heater takes no longer than 150 ms (see Fig.~\ref{fig::time_responses} a), b) and c) ), while the full reconfiguration of the optical chip requires up to two seconds due to thermal relaxation processes (see Fig.~\ref{fig::time_responses} e) and f) ). In addition, Fig. \ref{fig::time_responses} (d)  shows the stability of the chips' operation during a complete thermal power redistribution due to the switching from one transformation to another.

\section{Heater crosstalk \label{app:heater_crosstalk}}

\begin{figure*}[p]

    \includegraphics[width=2.0\columnwidth]{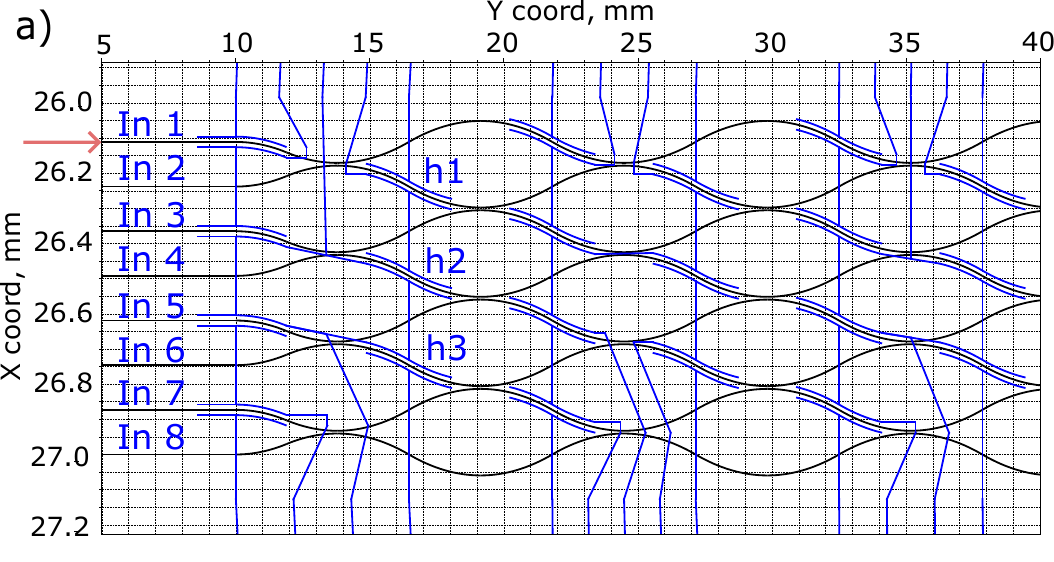}
    \includegraphics[width=2.1\columnwidth]{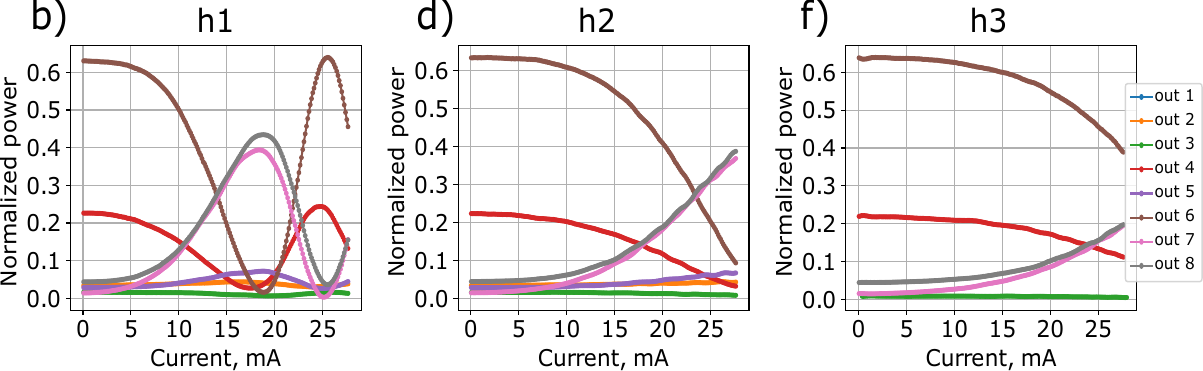}    
    \includegraphics[width=2.0\columnwidth]{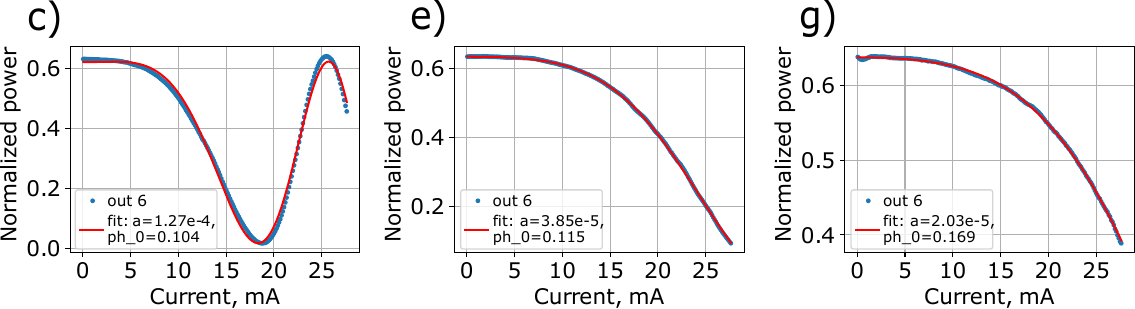}

\caption{\label{fig::heater_crosstal} Measured thermal crosstalk between heaters in the transverse direction. a) Crosstalk measurement protocol: light was injected into first input mode of the optical chip and electrical current was applied to a single heater (h1, h2 or h3, each with resistance of 480 $\Omega$) in range from 0 to 28 mA with 0.2 mA step. b) and c) Measured optical power during the current sweep through the heater h1 and the model fit of the data. The current corresponding to  $2\pi$ phaseshift is equal to 26.1 mA which yields the 0.33 W of dissipated heat. d) and e) Measured optical power during the current sweep through the heater h2 and the model fit of the data. f) and g) Measured optical power during the current sweep through the heater h3 and the model fit of the data. 
Phase shift induced by a heater is $\phi(x) = \alpha x^2 + \phi_0$, where $\phi$ - phase induced by the applied current $x$ and $\phi_0$ is the constant phase offset.}
\end{figure*}

We observe the crosstalk effect in the transverse direction in our PMI. We experimentally estimate the strength of a heating crosstalk between two heaters in a column as follows: coherent radiation from a diode laser is injected into the first input mode of the optical chip and electrical current is applied to a single heater --- h1, h2 or h3 (see Fig. \ref{fig::heater_crosstal} (a)) --- and the optical power from all outputs is logged down. Each heater h1, h2 and h3 has resistance of 480 $\Omega$). Current was applied in range from 0 to 28 mA with 0.2 mA step. Obtained optical power distributions between output modes were then fitted according to a linear power dependence law: 

    \begin{equation}\label{eq::power_vs_current}
        P_j = A - B \cos( \alpha x^2 + \phi_0 ),
    \end{equation}

where $P_j$ is the normalized optical power from $j$-th output port, $x$ - is the electrical current value, $\alpha$, $\phi_0$, $A$ and $B$ are the model parameters. The strength of the heater influence is characterized by the $\alpha$ parameter. Therefore, the relation between thermal influences can be estimated as the relation between the corresponding $\alpha_{hj}$ parameters.

The measured normalized output power distributions as the functions of the applied electrical current are shown in Fig. \ref{fig::heater_crosstal}. The output of the curve fitting algorithm is provided in Table \ref{tab::crosstalk_cmpr}, which shows that the crosstalks $\alpha_{h2}$ and $\alpha_{h3}$ are 3.26 and 6.35 times weaker than the direct h1 heater effect $\alpha_{h1}$.

The presence of the thermal crosstalks between neighbouring heaters may complicate the tuning of the PMI because they have to be accounted, for example, in a heater calibration procedure. However, the optimization routine automatically compensates the crosstalk effect and finds optimal phasehift configuration corresponding to the chosen transformation.

\begin{table}[h]
    \centering
    \begin{tabular}{ |c|c|c|c|c| } 
     \hline
     Heater & $\alpha$ & $\phi_0$ & A & B   \\ \hline
     h1 & 1.27e-4 &	0.104 & 0.68 & 0.30 \\ \hline
     h2 & 3.85e-5 &	0.115 & 0.68 & 0.32 \\ \hline
     h3 & 2.03e-5 &	0.169 & 0.68 & 0.32 \\ \hline
    
    \end{tabular}
    \caption{\label{tab::crosstalk_cmpr} Fitted parameters for three heaters in one vertical column (see Fig. \ref{fig::heater_crosstal}) characterizing the thermal crosstalk effect acting on the heater h1. The $\alpha$ parameter describes the strength of a heater. The induced phaseshift is $\phi(x) = \alpha x^2 + \phi_0$, where $\phi$ is phase induced by the current $x$ running through the heater and $\phi_0$ is a constant phase offset present even when the driving current is zero. }
\end{table}


\section{Power consumption \label{app:power_consumed}}

A typical phaseshift from applied electrical current dependence for a single heater with resistance of 480 $\Omega$ is shown in Fig. \ref{fig::heater_crosstal} a). The current corresponding to  $2\pi$ phaseshift is equal to 26.1 mA yielding the 0.33 W of the dissipated thermal energy. Thus, maximal electrical power applied to the optical chip can be estimated as 18.5 W. The thermoelectric element cooling power used for the PMI temperature stabilization is around 50 W. All the port-to-port optical mode switching configurations presented in the main text (Fig. \ref{fig::permut_result}) appeared to consume no more than 5 W of electrical power each. For instance, configurations with the power directed to the third output and to the sixth output displayed in Fig. \ref{fig::time_responses} (d-f) require 4.03 W and 3.71 W of electrical power, respectively.

\section{Numerical performance comparison with MZI-based PMI  \label{app:comparison_clement}}

We have conducted numerical simulations to compare the performance of the port-to-port optical mode switching between our PMI based on error-tolerant architecture \cite{Fldzhyan20} and the PMI composed of MZI blocks --- conventional universal PMI architecture \cite{Clements16}.  In our simulations we tested how inaccuracy in the directional couplers (comprising elements of both PMI architectures) will affect the ability to implement the port-to-port optical mode switching operation that we experimentally performed on our PMI.

We simulated the experiment of port-to-port optical mode switching, as it was held in practice
. The transmission coefficients of the DCs comprising the PMI were varied over a wide range to analyze the performance for a range of wavelengths. The ability to implement port-to-port optical mode switching was tested by running optimizations for each input and output port. If the phaseshift optimization run converged to the infidelity values lower than $10^{-3}$ for all $8$ permutations from the input port, then the PMI was considered capable of realizing the switching, otherwise it was considered incapable. If the PMI with the DC transmission coefficient $T$ (and with a particular input mode) satisfied the criterion, we put a marker on the plot. The results of the numerical simulations are shown in Fig. \ref{fig::comparison_permut}, which proves the higher tolerance to the transmission coefficients of the DCs in the BS-based error-tolerant architecture that we used in our PMI. The highlighted regions of switching capability are much wider for the BS-based architecture than for the conventional MZI-based PMI architecture \cite{Clements16}. The brightest illustrations of the error tolerance of our architecture are for the input ports 1, 2, 7, or 8, where the MZI-based PMI strictly supported one $T = 0.5$ DC transmission coefficient of perfectly balanced DC, while the BS-based PMI architecture has a wide range of available DC transmission coefficients.

\section{Infidelity convergence\label{app:converg}}
\begin{figure*}[t!]
\centering
\includegraphics[width=2.0\columnwidth]{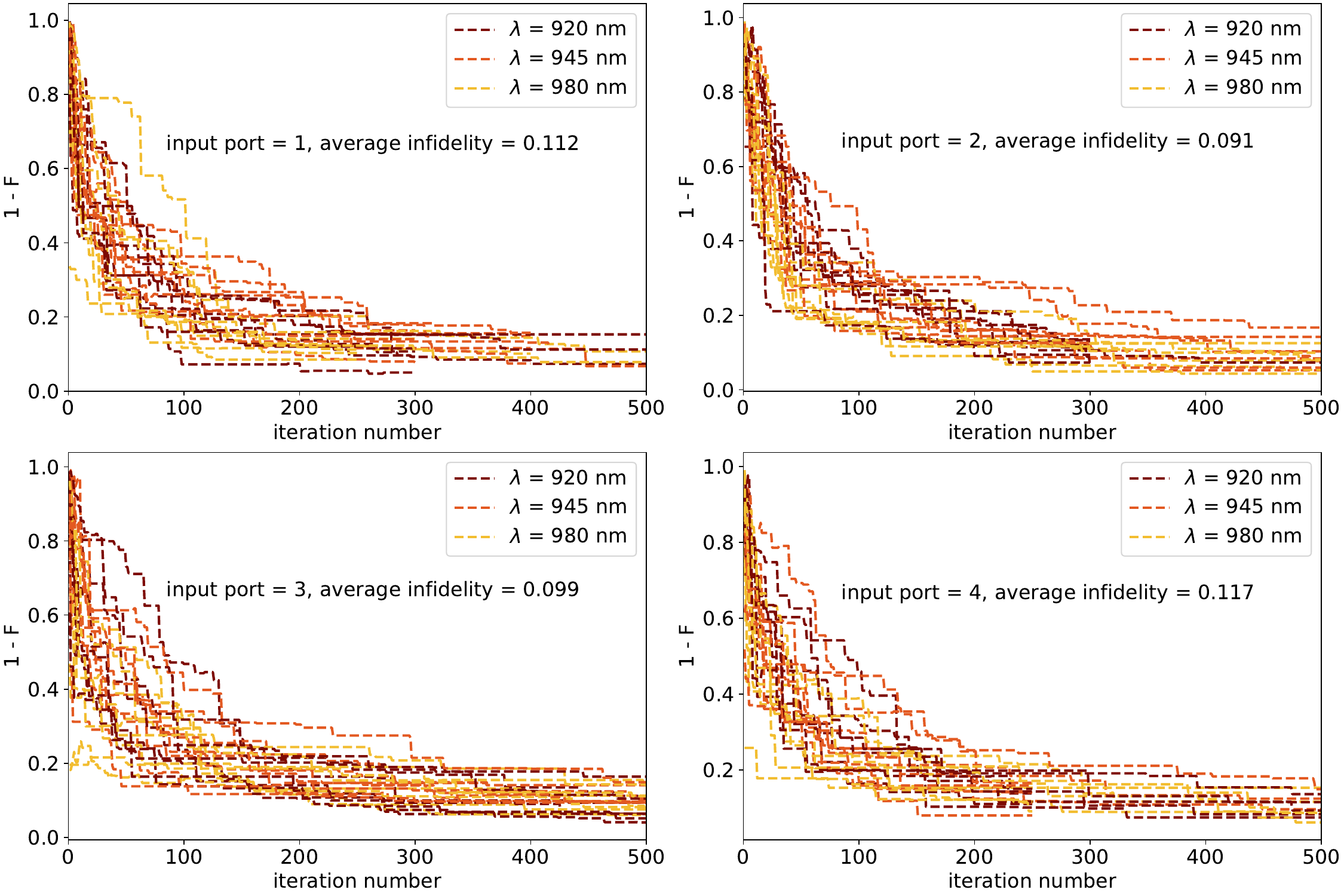}
\caption{\label{fig::converg} Convergence of the infidelity value. Each figure contains information about 24 optimizations. These are the results of phaseshift optimizations that minimize the infidelity between the target and measured output vectors. For each of the 4 first input ports, 8 optimization runs were performed to switch all radiation power to any of the output modes using laser light with 3 different wavelengths (920~nm, 945~nm, and 980~nm).}
\end{figure*}

The infidelity defined in Eq.~(\ref{eq::infidelity}) is used as a figure of merit of the distance between the measured ($X$) and the target ($Y$) vectors of power values at the output of the error-tolerant PMI. We minimize the infidelity value and establish  the vector of phaseshifts which implements the required transformation in the PMI. We limited the optimization algorithm to 500 iterations because longer optimization runs did not yield sufficiently better results. We conclude that the non-zero final infidelity values correspond to technological limitations of the interferometer fabrication. Figure~\ref{fig::converg} shows the infidelity convergence process for each of the four different input ports.

\begin{figure*}[t!]
\centering
\includegraphics[width=1.8\columnwidth]{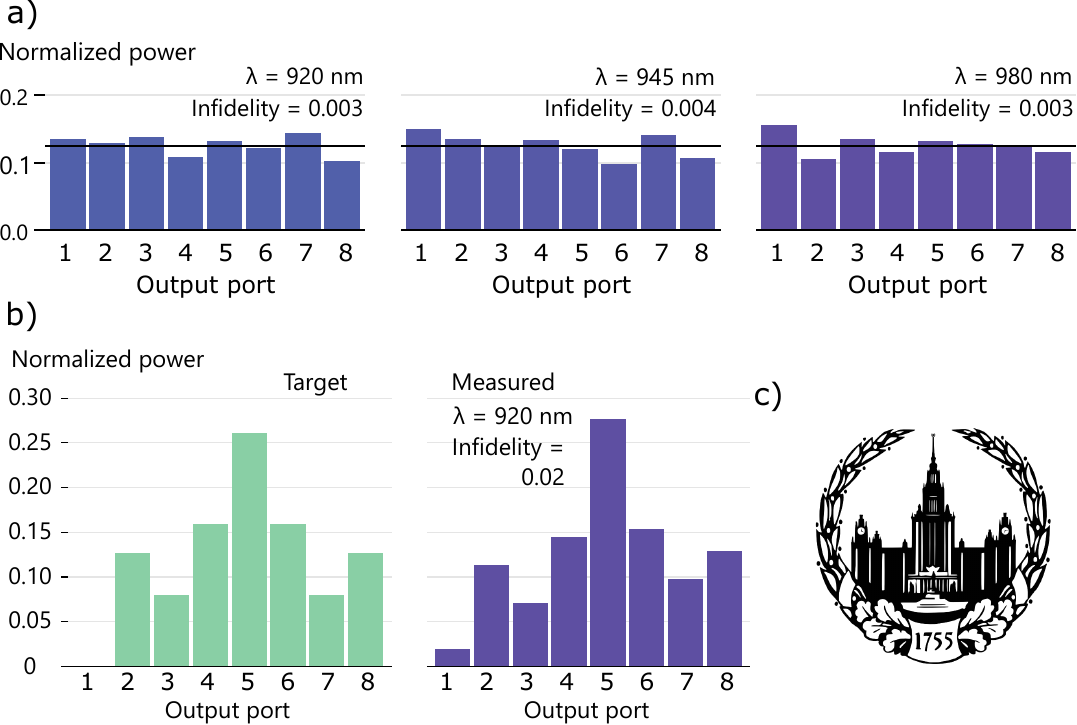}
\caption{\label{fig::distributions} Examples of achievable output power distributions of the studied optical chip with input radiation injected into the first port. a) Uniform power distributions obtained using three different laser wavelengths, b) power distributions replicating the shape of the main building of the Lomonosov Moscow State University, and c) the logo of the Lomonosov Moscow State University which illustrates the main building.}
\end{figure*}

The performance of the device was demonstrated by realizing a specific task --- broadband port-to-port optical switching, which implements transfer of the radiation energy from an input port to the chosen output port of the interferometer. However, according to the original proposal \cite{Fldzhyan20} the error-tolerant architecture guarantees the possibility to achieve an arbitrary power distribution at the output. As an example we prepared the uniform output power distribution using three different wavelengths, and the distributions that resemble the shape of the Lomonosov Moscow State University’s main building (see Fig. \ref{fig::distributions}). 

\section{Reconfigurable optical chip durability \label{app:durability}}

Durability of the reconfigurable optical chip is a vital parameter, as it shows how many cycles the optical chip can perform without malfunction of any of its’ components such as thermooptical phaseshifters. If even a single thermooptical phaseshifter breaks, the PMI will lose the universal reconfiguration feature resulting in a decay of optical chip performance, which will further decrease if the number of broken heaters grows.

For our PMI we observe both high level of performance and durability, which can be estimated by the total number of switching of each heater during the optimization runs conducted for obtaining the results presented in the main text (see Fig. \ref{fig::permut_result}).
Total number of optimization runs on the PMI was no less than 96 in order to obtain main result. Every optimization run on chip contained 500 iterations, where at each optimization step optical chip was reconfigured by applying new currents on all the 56 heaters, which leads to the overall $4\times8\times3\times500 = 48000$ switches of each heater on the chip and $48000 \times 56 \approx 2.7M$ heater switches in total.
In terms of operating time each optimization step was 10 seconds long, yielding to $48000 \times 10 (s) \approx 133$ hours of optical chip being instantly heated.  In fact, these estimations are lower bounds on the actual numbers of chips' operations, as, for example, some of the optimizations on chip were launched more than once. All 56 phaseshifters appeared to be undamaged afterwards, which evidences the high level of our PMI durability.

\end{document}